# ARTICLE

# Acoustic Probing of New Biomarkers for Rapid Sickle Cell Disease Screening

Nakul Sridhar,[a] Meiou Song,[a] Michael H.B. Stowell,[b] Kathryn L. Hassell,[c] and Xiaoyun Ding[*adef]

Sickle cell disease (SCD) remains a critical global health issue, with high child mortality in low-resource regions. Early screening and diagnosis is essential for improving health outcomes, but conventional screening methods are unsuitable for widespread use due to the high costs of laboratory equipment. There is an urgent need for portable, cost-effective, and user-friendly point-of-care tools that can quickly assess blood health. Here, we explore two new biomarkers enabled by acoustic probing for rapid SCD screening: cell membrane stability from measuring red blood cell (RBC) lysis temperature in whole blood, and plasma protein concentration from measuring relative protein precipitation in blood plasma. Both biomarkers effectively differentiate healthy HbAA samples from pre-/no transfusion HbSS samples with high accuracy. Additionally, the RBC lysis biomarker can distinguish post-transfusion exchange HbSS samples with a lower percentage of sickled cells, indicating the potential to initially screen for milder forms of SCD as well as sickle cell trait.

## Introduction

Sickle cell disease (SCD) is a severe inherited blood disorder that affects millions of people worldwide. In the most common form of the disease, a mutation in the HBB gene causes formation of abnormal sickle hemoglobin (HbS) that polymerizes under hypoxia, leading to the characteristic sickle shape of red blood cells (RBCs) and acute complications including severe anemia, pain crises, and organ damage[1–5]. In Africa, 50-90% of newborns with genotype HbSS, the most severe form of SCD, die before the age of 5[6]. Early screening and diagnosis can lead to timely intervention such as penicillin prophylaxis along with family education, which notably decreases the high mortality rate[7,8]. Laboratory tests, including high performance liquid chromatography (HPLC), hemoglobin electrophoresis, or isoelectric focusing (IEF), are considered as the gold standard, separating different Hb genotypes based on their different physical or chemical properties[9]. Additionally, Enzyme-linked immunoassays (ELISAs) employ antibodies to bind Hb variants, followed by enzymatic color changes using a substrate solution. An optical reader then quantifies the percentages of Hb variants[10,11]. These laboratory methods are accurate, sensitive, and reproducible; however, they necessitate expensive equipment, lengthy processing time, and expert interpretation, making them unsuitable for widespread use in resource-limited countries[12,13].

Recently, point-of-care (POC) technologies have emerged to address screening and diagnostic challenges in developing countries, improving cost-effectiveness, portability, and requiring minimal training to operate[12]. The sickle solubility test is a simple diagnostic method that is widely available and can visually confirm the presence of HbS in a blood sample. However, it cannot easily distinguish between more severe sickle cell anemia, milder forms of the disease or sickle cell trait (SCT)[14]. This limits the clinical usefulness of this test, especially in areas such as Sub-Saharan Africa where up to 30% of people may carry SCT[15,16]. The standard solubility test was modified by introducing a paper-based substrate that precipitates HbS onto paper and quantifies its percentage using imaging analysis[17–20]. Although it significantly improves differentiation between SCD and SCT, it has low specificity and may lack accuracy due to potential false negatives in newborns with high fetal hemoglobin (HbF) and low HbS levels.

Traditional laboratory tests have also been modified for use in point-of-care settings. Paper-based microchip electrophoresis, miniaturization of conventional electrophoresis, offers comparable results to gold standards but demands moderately complex laboratory processing[15,21]. Lateral flow assays utilize immunoassay mechanisms like ELISA but integrate all necessary components into one paper strip with high diagnostic sensitivity and specificity[22–25]. However, preparing testing strips is complex, and antibodies require strict storage conditions, making transport in extreme temperatures and humidities challenging.

As HbS-containing erythrocytes are denser than normal counterparts, Kumar et al. created an aqueous multiphase system (AMPSs) to separate RBCs into layers based on

a. Paul M. Rady Department of Mechanical Engineering, University of Colorado Boulder, CO, USA.
b. Department of Molecular, Cellular and Developmental Biology, University of Colorado Boulder, CO, USA
c. Colorado Sickle Cell Treatment and Research Center, School of Medicine, Anschutz Medical Campus, University of Colorado, Aurora, CO, USA.
d. Biomedical Engineering Program, University of Colorado Boulder, CO, USA.
e. Materials Science and Engineering Program, University of Colorado Boulder, CO, USA.
f. BioFrontiers Institute, University of Colorado, Boulder, CO, USA.
*Corresponding authors. E-mail: Xiaoyun.Ding@colorado.edu,
Supplementary Information available: [Fig. S1-S7, Movie S1-S2]. See DOI: 10.1039/x0xx00000x

densities[26,27], while Goreke et al. applied magnetic levitation principle to pull denser HbS-containing RBCs to the bottom of capillary and calculated the density range of HbS-containing RBCs and HbA-containing RBCs respectively for differentiation[28]. However, these methods were unable to screen for milder forms of the disease or SCT. Moreover, their reliance on extra equipment or complex procedures makes them less optimal for POC settings.

Microfluidic lab-on-a-chip devices have been widely utilized for disease screening due to their ability to facilitate cost-effective device fabrication, integrate operating systems into a compact form, and enable rapid tests[29,30]. Man et al. devised a gradient-microcapillaries chip to quantify occlusion blockage caused by the reduced deformability of HbS polymerization[31]. Alapan et al. utilized the endothelium-associated proteins, fibronectin and laminin, to gauge adhesive HbS-containing RBCs[32]. Liu et al. discerned discrepancies in electrical impedance signals between HbS-containing RBCs and normal RBCs, offering novel insights for SCD identification[33]. These approaches exhibited promising capabilities for accurately screening for SCD, but they still required complex device fabrication and expert execution of experiments.

The integration of surface acoustic wave (SAW) technology with microfluidics has found applications in diverse biological contexts, offering both high biocompatibility and precise fluid control[34–36]. Moreover, acoustic energy has been used to precisely heat and control on-chip temperatures[37–39]. This coupling of acoustic forces and acoustic-induced heating has enabled highly sensitive and label-free monitoring of protein-ligand interactions to identify drug candidates for protein targets[38]. This method also showed promise as a way to quickly distinguish healthy and SCD samples by measuring changes in thermal stability in sickled cells; however, it required preprocessed cell lysate and did not distinguish between SCD and SCT.

Here, we have discovered new biomarkers for SCD screening, enabled by acoustic probing: combining acoustic radiation force and acoustic heating. We apply surface acoustic waves (SAWs) to lyse RBCs and induce protein denaturation and aggregation on chip. Pathological changes within sickle cells will generate impact to cell integrity such as cell membrane integrity and protein thermal stability shift[5]. We track these impacts by introducing two new biomarkers—cell membrane stability by measuring the peak RBC lysis temperature point from whole blood samples, and plasma protein concentration by measuring relative protein precipitation from plasma samples—which both enable fast and easy initial screening for SCD. Our acoustic technique presents multiple advantages over current SCD screening tests: easy fabrication, minimal sample usage (less than 2 uL), rapid results in under 2 minutes, high accuracy, and no need for expertise or expensive equipment. Altogether, these acoustic-enabled biomarkers could pave the way for rapid SCD screening and overall blood health testing in POC settings.

## Experimental Methods

### Device fabrication

The SAW device consists of the lithium niobate (LiNbO3) substrate patterned with two identical IDTs on either side of PDMS channel. The IDTs were patterned onto a 500 μm thick, 76.2 mm, Y + 36° X-propagation LiNbO3 wafer using standard photolithography techniques. First, a layer of photoresist (S1813, Dow, USA) was spin-coated onto the wafer. The IDT structures were then patterned onto the wafer using UV light exposure and then developed using a photoresist developer (MF319, Dow, USA). Next, layers of chrome and gold (Cr/Au, 10/100 nm) were deposited using e-beam evaporation. Excess photoresist was removed using a lift-off process (Remover PG, Kayaku, Japan). Finally, the wafer was diced into individual devices each containing a pair of identical IDTs. Each individual IDT consists of 20 electrode pairs with a spacing of 50 μm and an aperture of 10 mm.

The PDMS chamber was made using a negative replica from an SU8 master mold. To fabricate the SU8 mold, an 80 μm thick layer of SU8 2025 photoresist (MicroChem, USA) was spin-coated on a 76.2 mm silicon wafer. The wafer was then patterned using standard optical lithography. To create PDMS replicas, the master mold surface was first modified by placing it in a silane vapor for 30 minutes. Then, a mixture of PDMS base and cross-linker (Sylgard 184, Dow Corning, USA) with a ratio of 10:1 w/w was poured onto the master mold and cured at 65 °C for one hour. Inlet and outlet holes with a 0.75 mm diameter, as well as a 0.35 mm diameter hole in the center of the chamber to measure the temperature, were punched using biopsy hole punchers. Finally, the PDMS chamber was bonded to the LiNbO3 substrate using oxygen plasma (PDC-001, Harrick Plasma, USA). To ensure a strong bond, the device was baked overnight at 65 °C.

### Samples and materials

Healthy human whole blood (WB) samples were purchased from Zen-Bio Inc (Research Triangle Park, NC, USA). Post-exchange waste HbSS sickle cell disease WB samples were obtained through the Colorado Sickle Cell Treatment and Research Center (Aurora, CO, USA). These samples contain a mixture of both HbSS cells and transfused normal HbAA RBCs. Pre- and no exchange HbSS sickle cell disease WB samples were obtained from the Sickle Cell Treatment and Research Center with informed consent under the Colorado Multiple Institutional Review Board approved protocol (Protocol #: 20-0505). These samples contain a higher percentage of HbSS RBCs as compared to the post-exchange waste blood samples. All WB samples were stored in K2 EDTA anticoagulant coated tubes (Vacutainer, BD, USA) at 4 °C. Human blood plasma was obtained by centrifuging the WB samples at 2000 ¬g for 10 min and stored at 4 °C. Some of the collected plasma from each sample was set aside for long-term storage at -20 °C and used for conventional protein concentration experiments. All WB samples were tested in the SAW device one day after collection. All plasma samples were tested in the SAW device within one week after collection. Samples were diluted in phosphate-buffered saline (PBS) (Thermo Fisher Science, USA) before experimentation. For all experiments, WB was diluted at a 1:10 WB:PBS (v/v) ratio and plasma was diluted at a 1:1 v/v ratio.

The following purified proteins and compounds were used: human γ-globulins (Millipore Sigma, USA), human hemoglobin (Millipore Sigma), human serum albumin (Millipore Sigma) and bilirubin (Millipore Sigma). For the elevated protein study, all proteins and compounds were diluted in PBS to create solutions at different concentrations and then mixed with plasma at a 1:1 v/v ratio. Bilirubin was first dissolved in dimethyl sulfoxide (DMSO) (Millipore Sigma) before PBS dilution, making sure that the final DMSO concentration was under 2% v/v to prevent damage to plasma proteins. For the BCA assay, bovine serum albumin (BSA) (Millipore Sigma) was used as a standard solution.

**Device operation**

An RF function generator (33500B, Keysight, USA) and power amplifier (403LA, E&I, USA) were used to input the signal into the IDTs via a pair of BCA cables. A set of PCB connectors transferred the signal from the BCA cables to the IDTs. The resonant frequency of the devices was identified using a network analyzer (E5061B, Keysight, USA). Depending on the individual device, the resonant frequency of the IDT transducers was identified as approximately 19.53-19.55 MHz.

Before each experiment, the PDMS channel was washed with concentrated bleach and water. Air bubbles were removed with ethanol, and PBS was manually preinjected into the inlet hole of the channel using a micropipette. For experiments, approximately 5 µL of diluted WB, plasma, or protein sample was manually injected into the channel, taking care that no air bubbles were left trapped inside. The SAW device was mounted on a custom-built 3D printed holder and placed on the stage of an inverted microscope (Eclipse Ti2, Nikon, Japan). The PDMS channel was imaged with a digital CMOS camera (Orca-Flash 4.0, Hamamatsu, Japan) in combination with the included imaging software (HCImage Live, Hamamatsu, Japan). All videos were recorded at 2.5 frames s-1 using a 10x objective under brightfield (lighting: 50% brightness, lowest aperture). Each captured image was 2048x2048 pixels. To ensure consistent measurements, the same channel area was imaged for all experiments. To monitor temperature inside the PDMS channel, a digital thermocouple connected to a DAQ system (NI, USA) and interfaced with LabVIEW (National Instruments, USA) was inserted in the middle of the channel and collected a measurement corresponding to every image frame.

**Image analysis**

All processing of videos and images was completed using ImageJ[40]. ROIs were initially traced using the rectangular selection tool. For WB images, a single ROI (1400x2000 pixels) across the entire channel area was used. For plasma images, ROIs (40x2000 pixels) were focused on all nodal lines when proteins aggregated such that background noise was limited and sensitivity was maximized (see Fig. S2A for example ROIs). To maintain consistency, the same ROI locations were selected for all experiments. Cell lysis and protein denaturation curves were generated using the *Plot z-axis profile* feature in ImageJ to calculate the average grayscale intensity in the ROI for each frame across the entire image stack. The peak cell lysis point in WB was quantified by matching the rightmost peak of the curve to its corresponding temperature (see **Fig. 1D**). The grayscale intensity change in plasma was calculated by subtracting the intensity value at 90 °C from the intensity at the first inflection point (see **Fig. 1E**).

**BCA protein assay**

25 µL of BSA standard was pipetted into microwells of a standard 96-microwell plate across a working range of 50-500 µg/mL. Plasma samples were diluted 150x, 300x, and 600x to match the working range and pipetted into the microwells. 200 µL of working reagent (Pierce BCA Protein Assay Kit, Thermo Fisher Scientific, USA) was added to each well and the plate was incubated at 37 °C for 30 minutes. Absorbance was then measured at 562 nm using a plate reader. Plasma concentration was calculated by linearly interpolating to the BSA standard absorbance values and multiplying by the corresponding dilution factor.

**Statistical analysis**

Statistical details are provided in the figure legends. All statistical analyses were completed using Excel (Microsoft, USA) and OriginPro (OriginLab, USA) software. For all tests of significance between data sets containing numerical means, a two-sided, unpaired Student's t-test was performed. Significance was determined as follows: *$P < 0.05$, **$P < 0.01$, ***$P < 0.001$. Receiver operator characteristic (ROC) curves were generated and the optimal threshold, area under curve, sensitivity, and specificity were extracted using the built-in ROC curve command in OriginPro. Diagnostic accuracy was calculated by dividing the total number of true positive and true negative tests by the total number of trials.

**SCD score calculation**

Cell lysis score was obtained by normalizing lysis temperature across entire WB sample subset from 0-1 using linear interpolation. Plasma precipitation score was obtained by normalizing grayscale intensity difference across entire plasma sample subset from 0-1 using linear interpolation. SCD score was obtained by adding cell lysis score and plasma precipitation score.Results and discussion

**Working principle of acoustic probing**

The schematic and working principle of the acoustic probing technique is shown in **Fig. 1**. The device consists of a rectangular polydimethylsiloxane (PDMS) channel filled with fluid bonded to a piezoelectric lithium niobate substrate. The 10 mm long channel has a height of 80 µm and a width of 1 mm. An identical pair of gold interdigital transducers (IDTs) are fabricated onto the lithium niobate substrate on either side of the channel. The device is placed in a custom-built holder fitted to a microscope stage and connected to BCA cables via a custom PCB connector (Fig. S1) Upon actuation via a radio frequency (RF) input (19.5 MHz), the IDTs produce SAWs which propagate along the surface of the substrate towards the PDMS channel. As the SAWs reach the fluid in the channel, the difference in velocity of acoustic propagation along the fluid versus the substrate causes leakage of some of the wave energy into the fluid. This energy imparted into the fluid gives rise to a rapid acoustic heating effect, which can be precisely controlled via the RF input power. Concurrently, interference caused by the SAWs propagating from the identical IDTs on either side of the channel results in the formation of a one-dimensional standing SAW field with areas of high-pressure nodes and low-pressure antinodes. Particles in the fluid are pushed towards and concentrated at the nodes.

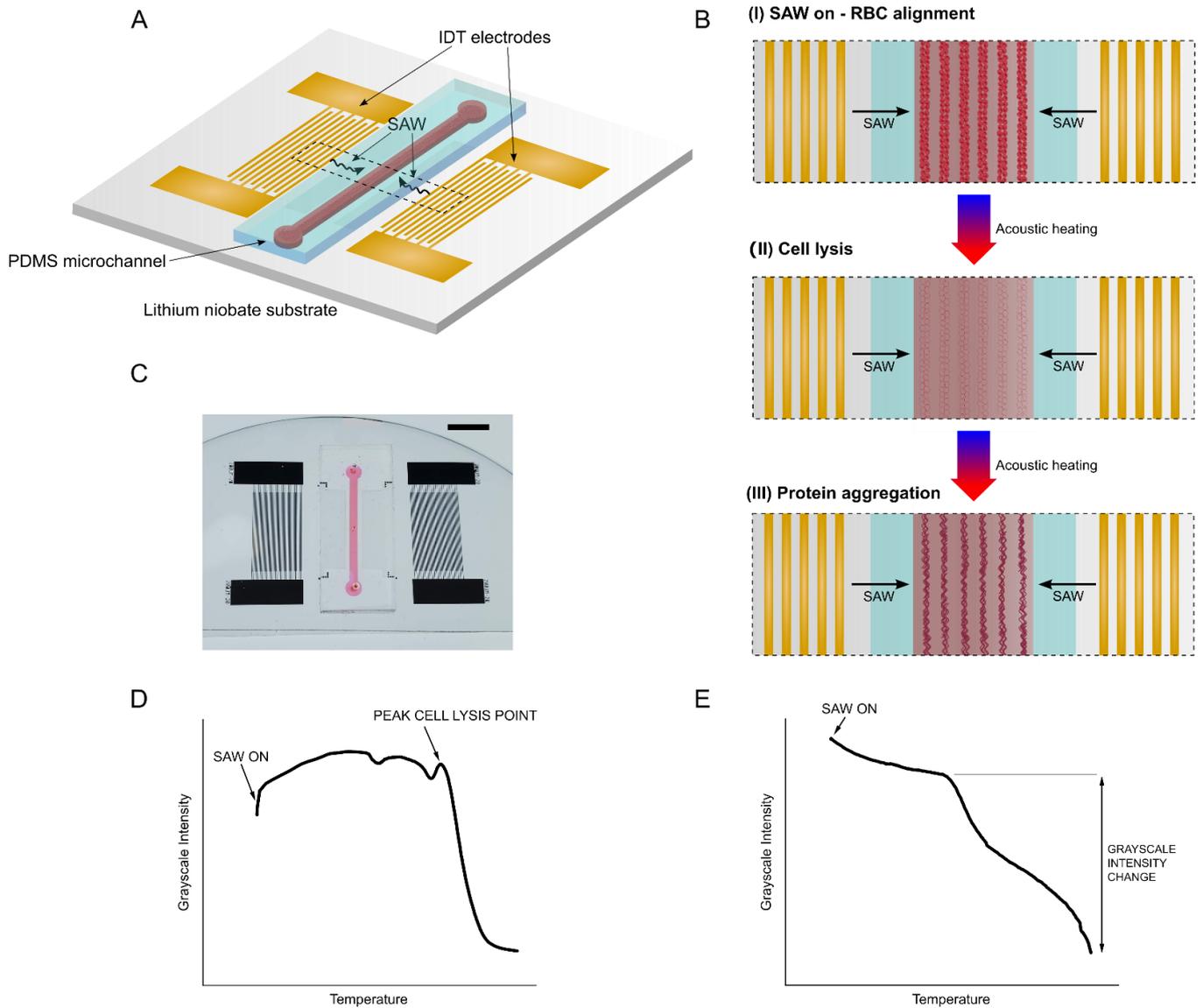

**Fig. 1. Working mechanism of SAW screening device using acoustic probing method. (A)** Schematic of SAW integrated device showing identical IDTs on either side of the PDMS channel fabricated onto a lithium niobate substrate. IDTs produce a standing SAW field inside the channel which induces acoustic patterning and heating. **(B)** Top view diagram illustrating the phases undergone by RBCs in whole blood the chamber upon SAW application: (i) RBC alignment along pressure nodes, (ii) RBC lysis, and (iii) protein denaturation, precipitation, and aggregation along pressure nodes. If plasma is placed in the channel instead, phases (i) and (ii) are skipped and the process proceeds directly to phase (iii). Cell and protein illustrations created with www.biorender.com. **(C)** Photo of example device. Scale bar: 5mm. **(D)** Cell lysis and protein precipitation curve generated from tracking average grayscale intensity as a function of temperature when using whole blood samples. Peak cell lysis point can be extracted from the curve. **(E)** Protein precipitation curve generated from tracking average grayscale intensity as a function of temperature when using plasma samples. Relative plasma protein precipitation can be extracted from the curve.

When the chamber is filled with whole blood, applying SAW immediately pushes the RBCs towards the pressure nodes, patterning and aligning them into concentrated lines (**Fig. 1B**). The fluid in the channel then rapidly heats up due to the acoustic heating effect, and eventually reaches a critical temperature where the cells start to lyse (**Fig. 1B**). As the temperature continues to increase, the released proteins from the cell start to denature, precipitate, and aggregate along the nodes (**Fig. 1B**). This entire process (Movie S1) can be visually tracked by measuring the average grayscale intensity of a selected region of interest (Fig. S2A) as a function of temperature and generating a cell lysis and protein precipitation curve (**Fig. 1D**). This curve follows a characteristic shape from which specific points corresponding to the critical steps in the lysis and precipitation process can be extracted, including the initial SAW induced patterning, the peak cell lysis point, and the onset of protein denaturation, precipitation, and aggregation. Specifically, we find that we can quantify the membrane stability of RBCs in a whole blood sample by measuring the peak cell lysis point.

If instead the channel is filled with pre-centrifuged blood plasma, a similar process as with whole blood unfolds, but bypasses the initial RBC patterning and lysis steps, since there are no cells contained in the plasma (Movie S2). By measuring the average grayscale intensity as a function of temperature, a protein precipitation curve can be generated (**Fig. 1E**). Then by calculating the difference between the measured grayscale intensity before and after the protein denaturation and precipitation process, we can quantify the protein concentration in a plasma sample by measuring the relative plasma protein precipitation. Using our acoustic probing technique, we demonstrate a proof-of-concept for using both novel biomarkers, RBC membrane stability and plasma protein concentration, for rapid screening of SCD.

**Characterizing the acoustic heating rate**

Previous work with traditional thermal shift assays has shown that the melting point of proteins is highly dependent on the applied heating rate[41,42]. Therefore, it was critical to make sure that the acoustic heating rate was precisely controllable and repeatable to achieve accurate results. We tracked the temperature in the channel for three different applied input powers at 20 MHz frequency and show negligible variability between individual tests for each power level (Fig. S3). RBC lysis and subsequent protein aggregation in whole blood samples started around 70 °C, and by 90 °C, the entire process was generally complete. Similarly, the protein denaturation and aggregation process in plasma started around 55 °C, and a majority was complete by 90 °C. Therefore, the 1.5 W input was required to reach high enough temperatures to ensure complete tracking. However, continuing to heat the chamber to 100 °C sometimes led to complications including bubble formation, evaporation, and the potential of device breakage from high thermal stress. Therefore, for all future experiments, we developed a standard protocol where 1.5 W input power was applied to increase the channel temperature from room temperature (21 °C) to 90 °C in approximately 110 seconds, after which the SAW input was turned off.

**RBC membrane stability is a biomarker for sickle cell disease**

Sickle cell disease (SCD) is an inherited disorder caused by the mutation of the β chain of hemoglobin (β-globin). The most common and severe form of the disease, sickle cell anemia, stems from the point mutation in position 6, where glutamic acid is substituted by valine, leading to the formation of sickle cell hemoglobin (HbS)[2,4]. When the HbS mutation occurs in two chromosomes, it is known as HbSS, where RBCs have no source of normal β-globin[43]. The presence of HbSS is characterized by numerous changes to the biomechanical properties of RBCs including mechanical fragility, poor deformability, and the typical sickling shape[5]. When the mutation occurs in only one chromosome, the remaining chromosome produces healthy β-globin, leading to nearly equal portions of healthy and sickled cells in individuals, i.e. HbAS. This is known as sickle cell trait (SCT), which is not associated with the typical clinical manifestations of sickle cell disease[44]. There are other less common genotypes as well, including HbSC and HbSβ-thalassemia among others. Standard electrophoresis measurements and other standard laboratory methods can quantify the exact percentages of each type of hemoglobin in a patient. For individuals with HbSS, HbS levels are usually greater than 80% with the remaining fraction being mostly fetal Hb F. For individuals with HbAS sickle cell trait, HbS levels are usually 40% or lower and approximately 60% health HbA[45].

We first investigated our ability to quantify RBC membrane stability using peak RBC lysis points from whole blood samples to distinguish between three different adult donor sets: healthy HbAA (control), and two sets of samples taken from various stages of the SCD transfusion exchange process or not on exchange[46]: post-exchange waste HbSS, and pre-/no exchange HbSS. Each donor group contained an approximate range of HbA and HbS proportions: control samples (HbA:100%, HbS:0%), post-exchange waste HbSS (HbA:70-75%, HbS:25-30%), pre-/no exchange HbSS (HbA:60-70%, HbS:30-40%). Although transfusions decrease the severity of the disease and changed the characteristic HbA and HbS percentages compared to those without any intervention, we were still able to test three distinct groups with different HbS levels to measure the diagnostic potential of our biomarker.

We used the developed SAW input protocol to track the average grayscale intensity of whole blood samples from each donor set as a function of temperature. A representative set of images with heightened contrast showing each phase in the lysis and precipitation process of a healthy HbAA control donor is shown in **Fig. 2A**. For all analysis, we used raw unchanged images (Fig. S2B). We then averaged the cell lysis and protein precipitation curves that were generated for each donor set (**Fig. 2B**). Initially RBCs were evenly dispersed throughout the channel. When SAW was turned on, the RBCs aligned to pressure nodes, which corresponded to a slight increase in the grayscale intensity. As the temperature in the chamber continued to increase, we tracked an initial peak in the grayscale intensity corresponding to a shift in the patterned alignment at approximately 55-60 °C. We believe this shift potentially occurs due to structural changes to the RBC membrane leading to changes in cell density[47]. We then identified a second sharp peak that began to form after approximately 70 °C. This was the critical point at which the majority of RBCs had been lysed – i.e. the peak cell lysis point. Interestingly, we found that the peak cell lysis point shows a rightward shift correlated with the disease severity of the donor group. On average, HbAA controls lyse at 70.9 °C, post exchange waste HbSS samples lyse at 73.2 °C, and pre-/no exchange HbSS samples lyse at 76.4 °C (**Fig. 2B**). Assessment of individual donor samples from each group showed the presence of three significantly different ranges of peak cell lysis corresponding to severity of disease (based on expected % of HbSS cells) (**Fig. 2C**), illustrating a difference in RBC stability between the difference groups and showcasing its use as a biomarker to screen for sickle cell disease.

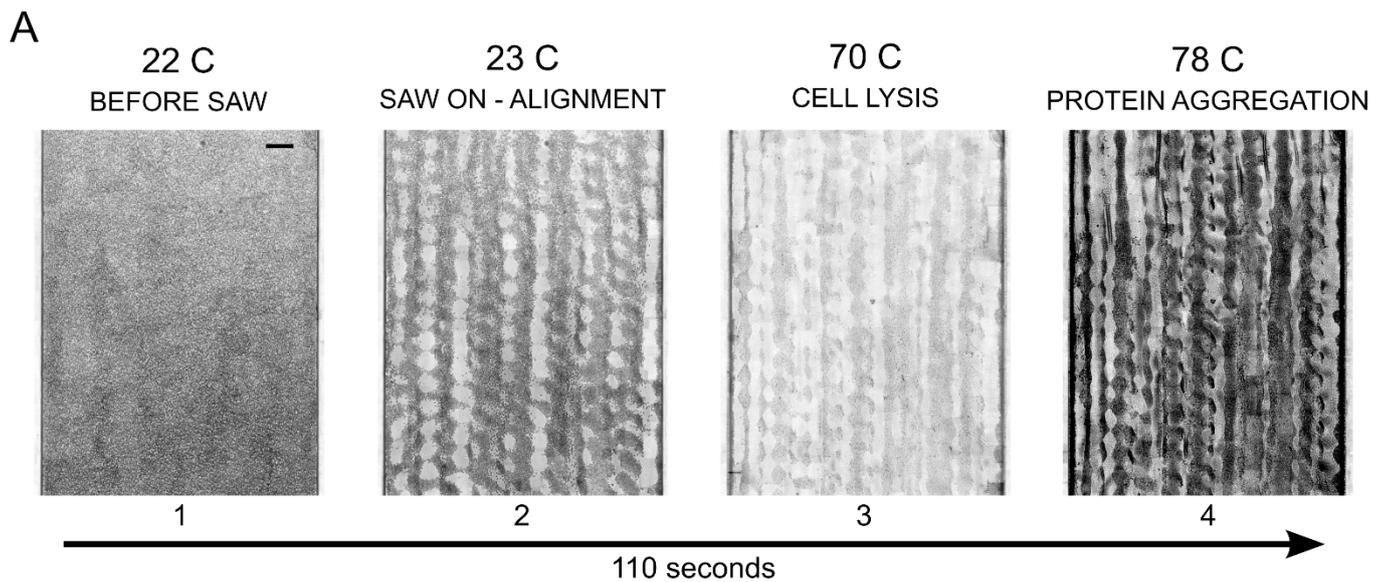

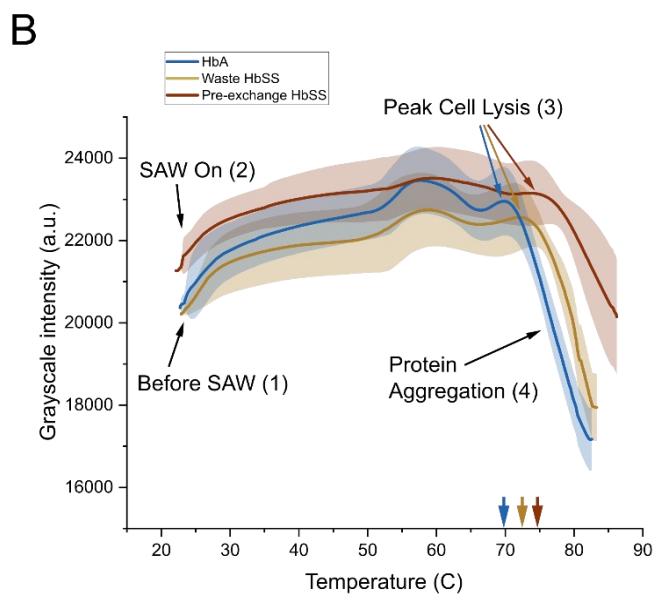
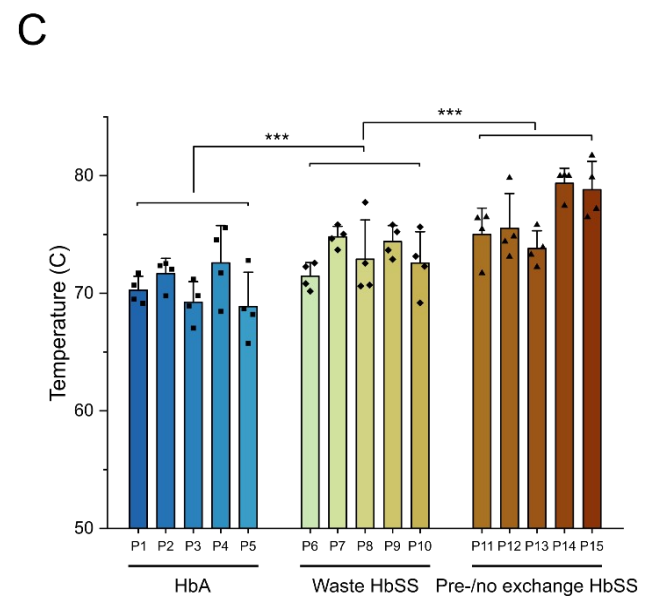

Fig. 2. RBC membrane stability as a diagnostic biomarker for sickle cell disease. (A) Representative images with enhanced contrast showing key phases of cell lysis and protein denaturation with corresponding temperature when SAW is applied to a whole blood sample. Images are taken from a single healthy HbAA donor (P3). Scale bar: 100 μm. Note: Raw images with unchanged contrast that were used for analysis are available in supplementary materials (see Fig. S2B). (B) Averages of cell lysis and protein denaturation curves comparing healthy HbAA (HbS:0%), post-exchange waste HbSS (HbS:25-30%), and pre-/no exchange HbSS donors (HbS:30-40%). Points on curve corresponding to representative images in (A) are labeled. The average peak cell lysis points for all three donor groups are shown (colored arrows). (C) Comparison of average peak cell lysis point measured for individual donors and sorted into the three donor groups as a measure of RBC membrane stability. Data are means ± s.d. For each group, 5 individual donors were tested with at least 4 independent trials per donor. Student t-test of independence was performed between all independent trials. ***P < 0.001; two-sided, unpaired t-test.

**Plasma protein concentration is a biomarker for sickle cell disease**

We then showcased the versatility of our acoustic probing method by replacing whole blood with blood plasma. Plasma is a commonly obtained diagnostic sample[48], and broadening the usable samples readily extends the utility of our technique. SCD is associated with an increase in total protein concentration in plasma[49], and measurements that can detect high levels of protein show promise as a diagnostic biomarker. Conventional methods to measure total protein concentration in plasma include colorimetric assays, such as the bicinchoninic acid (BCA) assay[50]. Specific proteins can be measured using ELISA[51,52]. However, neither are amenable to rapid POC screening due to high cost of reagents and equipment as well as lack of quality control, particularly in resource limited settings.

We centrifuged and extracted the supernatant plasma from the same whole blood samples that were used for the previous cell lysis tests. It is important to note that due to the dynamics of the transfusion exchange procedure where only RBCs are exchanged, plasma extracted from the post-exchange waste

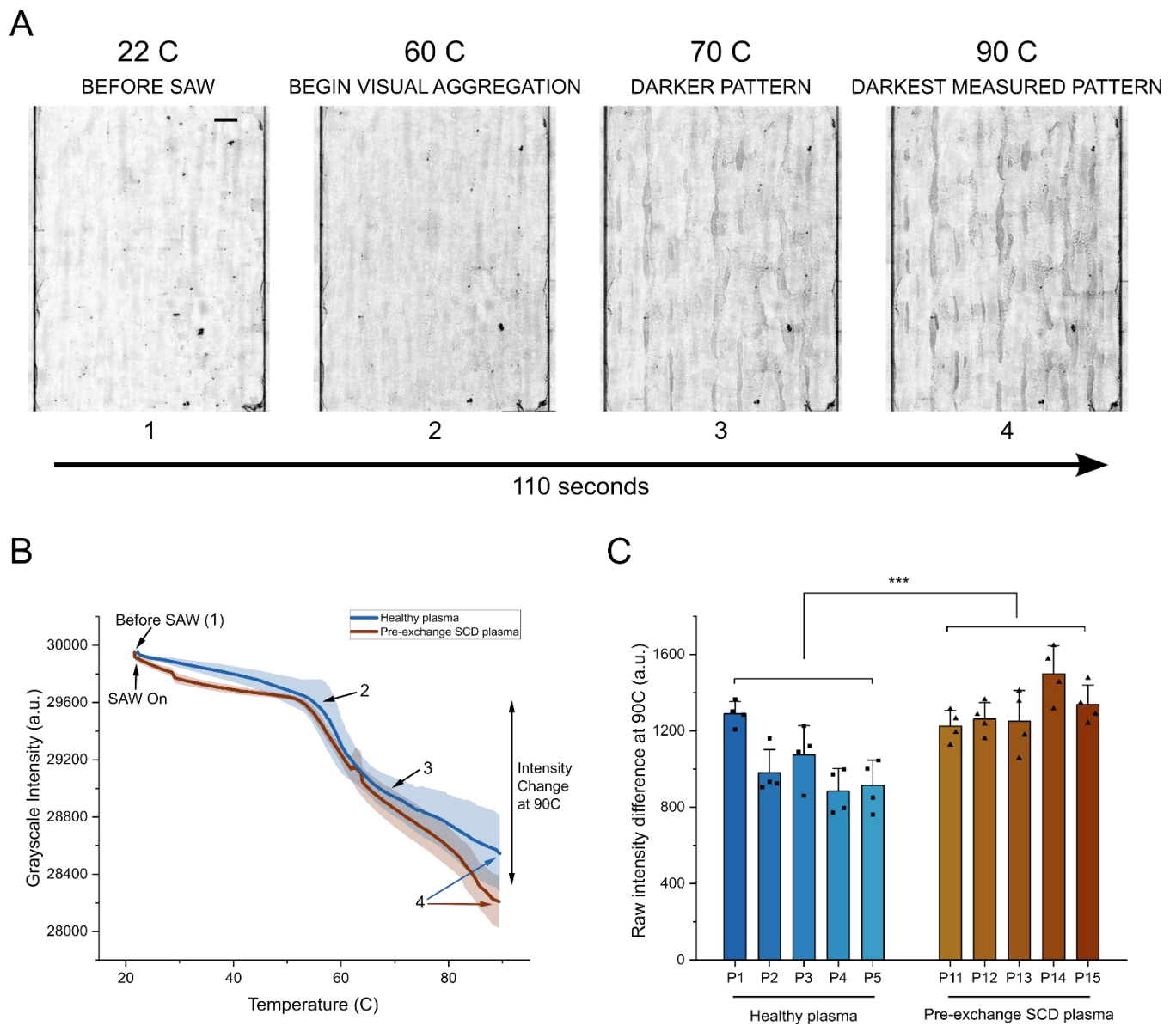

**Fig. 3. Elevated plasma protein concentration as a diagnostic biomarker for sickle cell disease. (A)** Representative images with enhanced contrast showing gradual darkening from increasing protein precipitation and aggregation at nodes with corresponding temperature when SAW was applied to a plasma sample. Images are of plasma that was taken from a single healthy HbAA donor (P3). Note: Raw images with unchanged contrast that were used for analysis are available in supplementary materials (see Fig. S2C). **(B)** Averages of protein precipitation curves comparing healthy plasma from HbAA donors, and SCD plasma from pre-/no exchange HbSS donors. Points on curve corresponding to representative images in (A) are labeled. **(C)** Comparison of relative plasma protein precipitation by calculating the average raw grayscale intensity difference measured at 90 C for individual donors and sorted into healthy and SCD groups as a measure of plasma protein concentration. Data are means ± s.d. For each group, 5 individual donors were tested with at least 4 independent trials per donor. Student t-test of independence was performed between all independent trials. ***P < 0.001; two-sided, unpaired t-test.

HbSS samples and the pre-/no exchange HbSS samples are functionally identical (Fig. S4), so we only focused on distinguishing between plasma taken from healthy HbAA controls (healthy plasma) and pre-/no exchange HbSS (pre-/no exchange SCD plasma). We used the same SAW protocol as before to heat the plasma samples such that they would denature, precipitate, and aggregate at the pressure nodes. A set of representative images with heightened contrast shows the gradual darkening patterns forming as more and more protein aggregates were visible (**Fig. 3A**). For analysis, raw images with no contrast alterations were used (Fig. S2C). We tracked and averaged the protein precipitation curves that were generated for each donor set and pinpointed the representative images corresponding intensity on the curve (**Fig. 3B**). Initially, the plasma was visually clear and showed the highest average intensity. Once SAW was turned on, the grayscale intensity slightly decreased even without any visual protein aggregation. This was due to changes in light refraction due to the presence of the acoustic pressure field. At approximately 55 °C, protein precipitates started to be visible, which could be seen on the curve as the first inflection point. Interestingly, we observed that after approximately 70 °C, a gap between the healthy plasma controls and the pre-/no exchange SCD samples could be seen. This gap increased at a higher rate after 80 °C until we stopped measurements at 90 °C (**Fig. 3B**). We quantified the average grayscale intensity change in plasma – i.e. the relative plasma protein precipitation by subtracting the measured intensity at 90 °C from the measured intensity at the first inflection point (~55 °C). Assessment of individual donor samples from both groups again shows a significant difference in the average intensity change (**Fig. 3C**). SCD samples showed an average change of 1309 a.u., while healthy controls showed an average change of 1029 a.u., a relative increase of 27% (280 a.u.) (**Fig. 3B**). This suggests that measuring higher intensity directly measures a relative increase in total protein concentration in the diseased sample. This straightforward use of our acoustic probing method to measure plasma protein concentration illustrates its potential use as a second biomarker to screen for SCD.

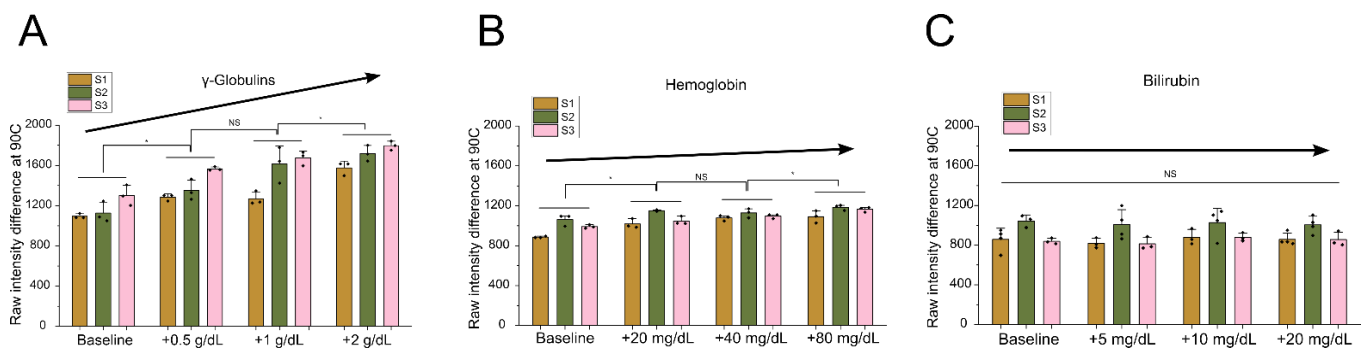

**Fig. 4. Assessing the impact that adding known concentrations of specific proteins or compounds has on the relative plasma protein precipitation signal. (A)** Characterizing the effect that purified γ-globulins added to healthy plasma samples have on the intensity change. Elevated γ-globulins concentration corresponds to a clear increase in the signal. **(B)** Characterizing the effect that purified hemoglobin added to healthy plasma samples has on the intensity change. Elevated hemoglobin concentration corresponds to a slight increase in the signal. **(C)** Characterizing the effect that purified bilirubin added to healthy plasma has on the intensity change. Elevated bilirubin plays no effect on the intensity signal. 3 individual healthy donor samples were tested at 4 different elevated protein concentration levels, with 3 independent trials for each measurement. *P < 0.05; two-sided, unpaired t-test. NS, not significant.

**Acoustic probing of protein concentrations in plasma**

Next, we further validated whether quantifying the relative plasma protein precipitation biomarker was an accurate measurement of the true protein concentration in a given plasma sample. To gain an understanding of how sensitive our measured intensity signal was to known changes in concentration, we tracked the intensity change for plasma samples for different dilution concentrations in PBS (Fig. S5A). Our results illustrated the clear correlation between the known plasma concentration and the measured relative protein precipitation using our acoustic method.

We then used a standard colorimetric BCA assay to measure the true total protein concentration levels in our tested samples to confirm that the SCD donors did indeed contain higher concentrations compared to the healthy donors. Our results showed that the SCD plasma samples showed a significant increase in measured concentration compared to healthy samples, with the SCD samples containing an average of 79.4 mg/mL and the healthy samples containing an average of 65.4 mg/mL, a relative increase of 21% (Fig. S6). This relative increase in concentration was quite similar to the relative plasma protein precipitation difference that we measured using our SAW device, again suggesting that our acoustic method was correctly measuring total protein concentration levels.

Subsequently, to identify if specific plasma proteins were having an outsized influence on the intensity change signal, we added known concentrations of individual proteins and compounds that are recognized as having increased levels in SCD. Considering our visual protein precipitation-based measurement, we expected that proteins found in the largest concentrations in plasma would have the most obvious effect on intensity change. The most abundant protein found in plasma is human serum albumin (HSA), which makes up more than half of the total protein present[53]. Most of the remaining fraction, approximately 40%, is made of the set of proteins known as globulins[54]. Previous studies have shown that SCD disease shows up to a 2x increase in globulin concentrations[49,55]. In particular, γ-globulins, a major class of immunoglobulins showed a marked concentration increase[56]. On the other hand, HSA has not shown any significant increase in SCD patients[49,55,56].

To estimate the effect that increased γ-globulins had on our grayscale intensity signal, we added known concentrations of purified γ-globulins to healthy plasma samples. We chose test concentrations based on expected γ-globulin values in SCD patients taken from previous studies[49,55,56]. We measured the grayscale intensity change for baseline healthy plasma samples and then compared when three increasing elevated concentrations of γ-globulins (+0.5, +1.0, +2.0 g/dL) were added to the same samples (**Fig. 4A**). The results clearly showed a significant increase in grayscale intensity change when increasing the γ-globulin concentration. Compared to the healthy plasma baseline, adding 0.5 g/dL increased intensity by an average of 226 a.u., 1.0 g/dL increased intensity by 345 a.u., and 2.0 g/dL increased intensity by 522 a.u. Interestingly, this relative increase was in the same range as what was previously measured in the SCD plasma samples compared to healthy plasma (an increase of 280 a.u.) (**Fig. 3B**). This suggests that γ-globulin concentration plays a major role in our measured plasma intensity change signal, indicating the presence of SCD.

We then tested the effect that free hemoglobin in plasma had on our intensity signal. Hemoglobin is contained within RBCs and is generally not present at significant levels in plasma[57]. However, SCD is known to cause chronic intravascular hemolysis, which releases the cellular contents and results in significantly elevated plasma hemoglobin levels[58]. Similar to the globulins test, we added known concentrations of purified hemoglobin to healthy plasma samples. We chose test concentrations of free hemoglobin based on expected values measured in previous studies[58]. As before, we measured the grayscale intensity change for baseline healthy samples and compared it with three elevated hemoglobin concentrations (+20, +40, +80 mg/dL) (**Fig. 4B**). It is important to note that although these test concentrations were significantly elevated compared to healthy samples, they were an order of magnitude lower than in the globulins test. The results showed a slight increase in grayscale intensity change as the hemoglobin concentration was increased. Compared to the healthy plasma baseline, adding 20 mg/dL hemoglobin increased intensity an average of 83 a.u., 40 mg/dL increased intensity by 121 a.u., and 80 mg/dL increased intensity by 166 a.u. This suggests that elevated hemoglobin in plasma potentially plays a small role in our measured SCD signal, but it is far less important than globulin.

**Table 1. Select diagnostic characteristics extracted from ROC curves.** Both RBC membrane stability and plasma protein concentration characteristics were calculated using optimal threshold values.

| RBC membrane stability | AUC | Sensitivity | Specificity | Accuracy |
| --- | --- | --- | --- | --- |
| HbAA vs Pre-/no exchange HbSS | 1.00 | 1.00 | 1.00 | 1.00 |
| HbAA vs Waste HbSS | 0.88 | 0.80 | 0.80 | 0.80 |
| Waste HbSS vs Pre-/no exchange HbSS | 0.92 | 0.80 | 0.80 | 0.80 |
| **Plasma protein concentration** | **AUC** | **Sensitivity** | **Specificity** | **Accuracy** |
| Healthy plasma vs SCD plasma | 0.88 | 0.80 | 0.80 | 0.90 |

To explore whether non-protein compounds played any role in our grayscale intensity measurement, we tested the effect that elevated bilirubin had on the signal. Bilirubin is a yellowish pigment compound that occurs during the breakdown of hemoglobin[59]. Elevated bilirubin is a common manifestation in SCD patients, stemming from chronic hemolysis[60]. As before, we added known concentrations of purified bilirubin to healthy plasma samples, choosing test concentrations based on expected values measured in previous studies. We measured the grayscale intensity change for the baseline healthy plasma samples and compared it with three elevated concentrations of bilirubin (+5, +10, +20 mg/dL) (**Fig. 4C**). These tests showed no increase in the grayscale intensity change when bilirubin was added to the healthy plasma sample. This confirmed that our intensity signal was dependent specifically on the protein denaturation, precipitation, and aggregation process, rather than the presence of increased pigment in the sample.

We then investigated if the denaturation of specific proteins could be mapped to the general protein denaturation curve that we measured for plasma. We first plotted the denaturation curves of the most abundant proteins found in plasma, specifically human serum albumin (HSA) and globulins, using estimated physiological concentrations (Fig. S7). To simplify the study, we used purified γ-globulins as an estimation for globulins overall. We also plotted the denaturation curve of purified hemoglobin, which may have a small effect on the general denaturation curve as well (Fig. S7). Both the γ-globulins and HSA showed denaturation starting to occur after 75 °C. This corresponds with a major part of the intensity difference between healthy plasma and SCD plasma, which also occurs after 75 °C. Previous work has not described significantly increased HSA in SCD patients[49,55,56], so our result again points to γ-globulins playing a key part in our SCD signal. For hemoglobin, although visual aggregation could be starting at approximately 55 °C, the change in grayscale intensity was negligible compared to both HSA and γ-globulins, again suggesting it is not a key part of our measured signal.

**ROC curves and combined SCD score quantify diagnostic capability of these new biomarkers**

Finally, we quantitatively characterized the diagnostic capability of both biomarkers using our acoustic probing method by plotted receiver operating characteristic (ROC) curves comparing samples for all donor groups (n=5 per group). ROC curves plot the true positive rate (sensitivity) against the false positive rate (1-sensitivity) for a full range of threshold values. For the RBC membrane stability biomarker using the peak cell lysis point, we plotted three ROC curves comparing the three different groups (**Fig. 5A**), and for the elevated plasma protein concentration biomarker using relative protein precipitation, we plotted one ROC curve comparing two groups (**Fig. 5B**). We then calculated the area under curve (AUC) comparing each group. By measuring RBC membrane stability, we could perfectly distinguish between HbAA and pre-/no exchange HbSS with an AUC of 1.00. Our method also showed a strong ability to distinguish between HbAA and waste HbSS, as well as waste HbSS and pre-/no exchange HbSS, with AUCs of 0.88 and 0.92 respectively. We also plotted the optimum threshold values, from which we extracted the diagnostic sensitivity, specificity, and accuracy for each comparison (**Table 1**), which shows similar performance in distinguishing HbAA and HbSS compared to other recently developed POC screening methods [26–28], while also showing strong potential for screening for less severe forms of disease or SCT, which could be approximated by our post-exchange waste HbSS sample.

We also assessed whether combining both biomarkers would further improve the diagnostic capability of our acoustic technique. We had RBC membrane stability and plasma protein concentration information for two donor groups: HbAA/healthy plasma and pre-/no exchange HbSS/SCD plasma. For all donors in both groups, we normalized the RBC membrane stability and plasma protein concentration each to a 0-1 range, and then added the two normalized values together to create a combined SCD score ranged between 0-2 (**Fig. 5C**). Scores closer to 0 correspond to healthier samples, while scores closer to 2 suggest the presence of disease. Similar to the RBC membrane stability screen, this combined scoring method also provided a 100% accurate differentiation between the healthy and SCD group. Moreover, by using the combined SCD score, we corrected for individual samples that provided a misdiagnosis when using a single screening method (for example, the outlier P1-healthy plasma sample – see **Fig. 3C**), suggesting a more robust screen that is less affected by individual outliers. This improvement after combining scores was also seen when comparing the healthy group to the post-exchange waste group (Fig. S7). We did not compare the post-exchange waste SCD samples and pre-exchange SCD samples because the plasma samples were functionally identical due to the transfusion process (Fig. S4).

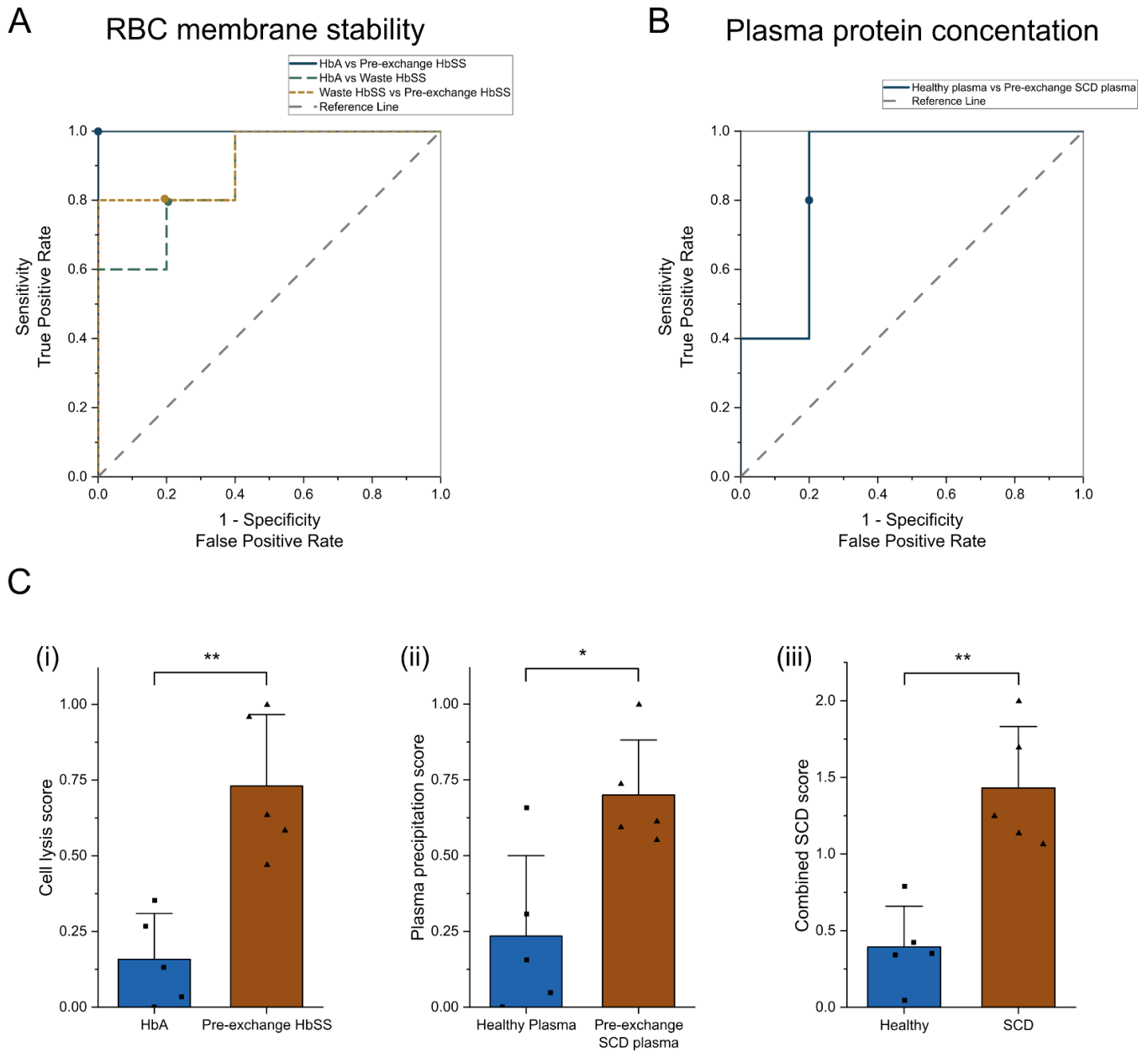

**Fig. 5. Receiver operator characteristic (ROC) curves quantifying the diagnostic capability of the new acoustic enabled biomarkers. (A)** ROC curves evaluating diagnostic performance of RBC membrane stability measurement in whole blood comparing healthy HbAA, post-exchange waste HbSS, and pre-/no exchange HbSS. **(B)** ROC curve evaluating diagnostic performance of plasma protein concentration measurement comparing healthy and SCD plasma. Optimal threshold for all curves is labeled. Compared to random guess reference line (dashed gray), all curves are close to the top left-hand corner demonstrating good diagnostic performance. **(C)** Combined diagnostic by generating normalized score (0-1) for both (i) RBC membrane stability and (ii) plasma protein concentration and adding together to produce (iii) combined SCD score (0-2). Pink squares – cell lysis and plasma precipitation score from donor P1. Green square – combined SCD score for donor P1 suppresses outlier. For each group, 5 individual donors were tested with at least 4 independent trials per donor. Student t-test of independence was performed between individual donors. **P < 0.01 and *P < 0.05; two-sided, unpaired t-test.

## Conclusions

Sickle cell disease continues to be a critical global health concern, with high childhood mortality rates affecting low-income regions. The highest priority for solving this issue is providing more widespread and affordable access to early screening and diagnosis, particularly for newborns. This allows for earlier interventions which can dramatically improve health outcomes. Conventional testing methods such as HPLC and electrophoresis require bulky lab equipment and/or high capital cost, making them unfeasible for widespread use in areas where resources are limited. Therefore, the lack of inexpensive and easy to use testing methods continues to severely limit efforts to implement more widespread newborn screening programs[15]. To address these challenges, new POC tests have recently been developed[12,13]. However, they still suffer from potential drawbacks to their cost effectiveness and usability in the field. Our new acoustic probing-enabled biomarkers provide an exciting screening technique with the potential to address these challenges. The acoustic device can handle unpurified whole blood and plasma samples, requiring no special reagents or labeling other than standard saline for dilution. The testing itself is rapid (under 2 minutes), requires low sample volume (2 μL), and the visual grayscale intensity-based tracking is straightforward and amenable to automated analysis. The power requirements for the device are quite low, and we envision packaging the chip into a smart phone integrated system capable of providing the input power, imaging, and analysis all in an easy to use, portable format. Sample pre-processing such as microcentrifugation to separate out plasma and RBCs can also be integrated on-chip[30]. The simple design of the chip lends itself to straightforward fabrication using standard photolithography techniques, which can be easily scaled up to mass production levels.

We use acoustics to introduce two novel biomarkers, RBC membrane stability and plasma protein concentration, both of which not only can effectively screen for the presence of SCD, but also show promise for more generalized use as well. In other studies using cells, lysis has generally simply been induced as a preprocessing step to release the internal cellular contents for analysis[61]. Here, our SCD screening results show that the precise temperature at which cell lysis occurs itself provides valuable information about the health of cells. Our simple lysis mechanism via rapid acoustic heating provides a generalized platform for further studies to potentially connect the lysis temperature with both the biomechanical integrity of cell membranes and cellular function for other diseases.

Accurate measurement of protein concentrations is a critical step in numerous biological studies, including characterization of cellular function[62] and drug-ligand binding[63]. Standard assays for protein concentration measurements generally require specific buffer reagents and spectrophotometry equipment[50]. By tracking the precipitation and aggregation of proteins under acoustic heating, our platform facilitates a novel and straightforward method to quantify the protein concentration in a sample without the need for specific assay kits and complicated reagent preparation, especially for highly concentrated solutions.

For POC use, a successful screening method relies on its cost-effectiveness, versatility, and ease of use. Our acoustic probing method introduces two new biomarkers, RBC membrane stability and plasma protein concentration, which we demonstrate can successfully screen for sickle cell disease. Our initial validation study shows strong diagnostic performance, with the ability to distinguish between healthy (HbAA), mild (post-exchange waste HbSS) and more severe (pre-/no exchange HbSS) donor groups. For widespread initial pre-screenings where large groups of people need to be tested cheaply and quickly, full quantitative information about specific Hb percentages and the exact variant of disease may not be the most critical information that people initially require. Rather, an initial overall indication of blood health can inform those individuals for whom more extensive lab-based testing is needed. We believe that our acoustic method is ideal for use in such initial pre-screenings.

We would next like to further validate our technique with larger clinical sample sizes with more precise pre-analytical quantification of HbS-containing cells, such as HbS% and clinical blood count, to confirm the diagnostic capability of the device, as well as investigate its effectiveness at distinguishing between other SCD genotypes, such as HbSC and HbSβ-thalassemia. Additionally, for future iterations of the device, we would like to replace our current microscope-based imaging setup with a small integrated camera system that is more amenable for POC use. Further automation of the SAW input system and temperature feedback control for increased reliability is also being developed.

Moreover, this acoustic method also shows potential for broader applicability to study the role that cellular thermal stability and protein concentration levels may play in other blood-based disorders. Specifically, it should be noted that alterations in red blood cell stability, including changes in deformability and membrane fluidity, are not exclusive to sickle cell disease. Other hematological conditions are known to significantly impact RBC biomechanical properties, including genetic blood disorders such as hereditary spherocytosis and hereditary elliptocytosis[64], parasitic infections like malaria[65], and metabolic diseases such as diabetes mellitus[66]. While this study focused specifically on sickle cell disease samples to validate our acoustic probing method, our results demonstrate the technique's ability to distinguish diseased RBCs from healthy cells based on membrane stability differences. This suggests broader applicability of our surface acoustic wave-enabled approach as a rapid initial screening platform for various blood-related disorders, warranting future investigations into its diagnostic potential across multiple hematological conditions that affect red blood cell integrity. Overall, our acoustically enabled biomarkers show promise for cheap, rapid, and accurate hematology in POC settings.

## Author contributions

Research Design: NS, XD
Experiments: NS, MS
Supervision: KLH, XD
Funding Acquisition: XD
Discussion: NS, MS, MHBS, KLH, XD
Analysis: NS, MS, MHBS, KLH, XD
Writing: NS, MS, MHBS, KLH, XD

## Conflicts of interest

A patent application based on this work is pending.

## Data availability

The data supporting this article have been included as part of the paper and the Supplementary Materials.

## Acknowledgements

This research was supported by the NIH MIRA Award 1R35GM142817 and the University of Colorado Lab Venture Challenge Seed Grant. We thank Julie McAfee for helping with the sample handling in the Colorado Sickle Cell Treatment and Research Center. The microfluidic devices were fabricated in JILA clean room at University of Colorado Boulder.

## Notes and references